\documentclass[a4paper]{spie}  
\usepackage[]{graphicx}
\usepackage[]{amsmath}
\usepackage[labelsep=period]{caption}
\usepackage{graphicx}
\usepackage{subcaption}
\usepackage{adjustbox}
\usepackage[labelfont=bf,labelsep=period]{caption}
\usepackage{multirow}
\usepackage{booktabs}
\usepackage{bm}

\DeclareMathAlphabet\mathbfcal{OMS}{cmsy}{b}{n}

\title{Organ Segmentation From Full-size CT Images Using Memory-Efficient FCN} 

\author{
Chenglong Wang\supit{a}, Masahiro Oda\supit{b}, Kensaku Mori\supit{b,c}
\skiplinehalf
\supit{a}Graduate School of Information Science, Nagoya University, \\Furo-cho, Chikusa-ku, Nagoya, Aichi, 464-8601, Japan; \\
\supit{b}Graduate School of Informatics, Nagoya University, Japan; \\
\supit{c}Research Center for Medical Bigdata, National Institute of Informatics, Japan 
}

\begin{document} 
\maketitle 

\begin{abstract}
    In this work, we present a memory-efficient fully convolutional network (FCN) incorporated with several memory-optimized techniques to reduce the run-time GPU memory demand during training phase. In medical image segmentation tasks, subvolume cropping has become a common preprocessing. Subvolumes (or small patch volumes) were cropped to reduce GPU memory demand. However, small patch volumes capture less spatial context that leads to lower accuracy. As a pilot study, the purpose of this work is to propose a memory-efficient FCN which enables us to train the model on full size CT image directly without subvolume cropping, while maintaining the segmentation accuracy. We optimize our network from both architecture and implementation. With the development of computing hardware, such as graphics processing unit (GPU) and tensor processing unit (TPU), now deep learning applications is able to train networks with large datasets within acceptable time. Among these applications, semantic segmentation using fully convolutional network (FCN) also has gained a significant improvement against traditional image processing approaches in both computer vision and medical image processing fields. However, unlike general color images used in computer vision tasks, medical images have larger scales than color images such as 3D computed tomography (CT) images, micro CT images, and histopathological images. For training these medical images, the large demand of computing resource become a severe problem. In this paper, we present a memory-efficient FCN to tackle the high GPU memory demand challenge in organ segmentation problem from clinical CT images. The experimental results demonstrated that our GPU memory demand is about 40\% of baseline architecture, parameter amount is about 30\% of the baseline. 
\end{abstract}

\keywords{Organ segmentation, memory efficient, FCN}

\section{Introduction}
Nowadays, deep-learning based approaches have gained a significant improvement against traditional image processing approaches in both computer vision and medical image processing fields. One main factor that makes deep learning techniques practical is the rapid development of computing hardware. With the high-performance computing hardware, the training time can be shortened to days or even hours. However, problem become more difficult in medical image processing fields, unlike general color images used in computer vision tasks, medical images have larger scales than color images such as 3D computed tomography (CT) images, micro CT images, and histopathological images. To fit the GPU memory limitations, one common processing is to feed to the cropped subvolume to neural networks. However, the cropped subvolumes have smaller field of view (FOV) which leads to lower accuracy. How to take the balance between the subvolume size and batch size has become a common problem for many medical processing tasks. A lot of research has been done on neural network pruning in the literature. He \textit{et al.} used a combination of a LASSO regression and least square reconstruction method to reduce the number of channels in each layer \cite{he2017channel}. Liu \textit{et al.} pruned the channels by learning channel-wise scaling factors \cite{liu2017learning}. These channel pruning methods can decrease the run-time memory foot-print and accelerate training process. However, one disadvantage of channel pruning methods is that pre-training process is necessary, \textit{i.e.} to learn the unimportant channels, a full pre-training of original architecture is indispensable. Instead of pruning channels, weight pruning is a technique to reduce the number of weights \cite{srinivas2017training,han2015deep}. However, this technique mainly contribute to generate a smaller model which makes it possible to be stored in small storage. Another well-known network compression technique is weight quantization \cite{courbariaux2016binarized,jacob2018quantization,tang2017train}. However, similar to weight pruning techniques, due to the hardware limitations, quantization techniques can neither save run-time memory nor training time. The weights will be restored to original precise in GPUs.

In this paper, we aim to reduce the run-time memory usage to fit the large GPU memory demand. We tackle this problem from two aspects: architecture and implementation. The main contributions of this work can be summarized as: 1) a memory-efficient fully convolutional network (FCN) is proposed incorporated with several memory-optimized techniques to reduce the run-time GPU memory foot-print. 2) the presented memory-optimized techniques is easy-implemented and flexible. They can be easily incorporated to other networks.

\begin{figure}[tb]
    \centering
    \includegraphics[width=0.95\linewidth]{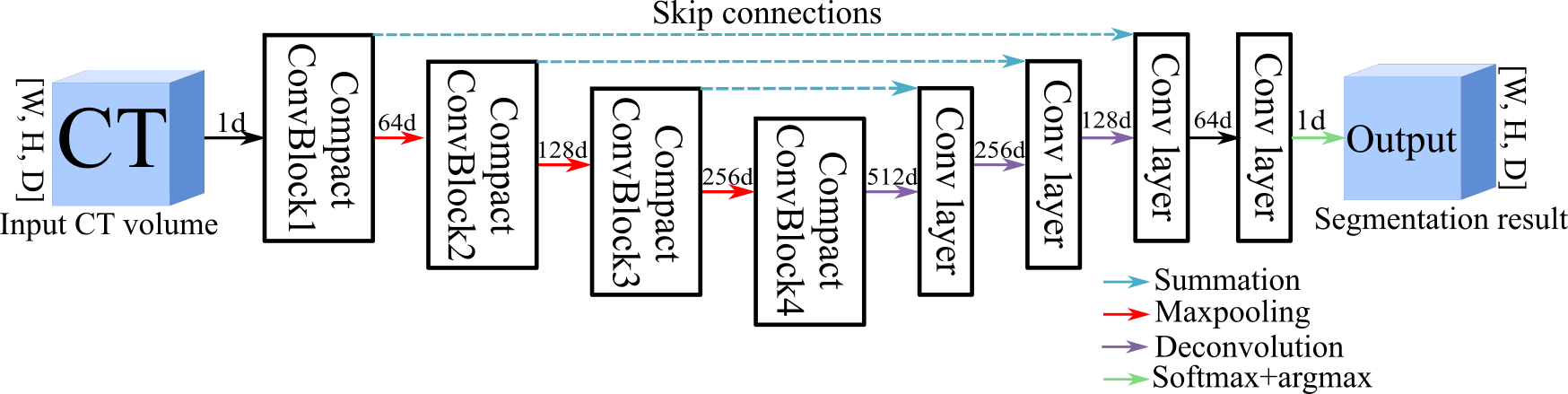}
    \caption{Overview of our backbone U-Net like architecture. Architecture of compact conv blocks are shown in Fig. \ref{fig:convblock}.}
    \label{fig:overview}
\end{figure}
\section{Method}
\subsection{Overview}
The motivation of this work is to reduce the GPU memory demand in training phase. We optimize our FCN from two aspects including architecture and implementation. From architecture aspect, we use bottleneck \cite{he2016deep} and depth-wise separable convolutions \cite{howard2017mobilenets} to reduce the number of parameters. From implementation aspect, we use both in-place activated batch normalization layer (Inplace-ABN) \cite{rota2018place} and mixed-precision (MP) training \cite{micikevicius2018mixed} technique to reduce the internal buffer size allocated in GPU memory. Overview architecture of our proposed FCN is illustrated in Fig. \ref{fig:overview}.

\subsection{Compact conv block}
Conventional U-Net architecture used two convolutional layers to extract features at each scale level \cite{cciccek20163d}. In this work, we first use a bottleneck structure to reduce the number of features and weight parameters. Bottleneck structure is introduced in ResNet \cite{he2016deep}. Bottleneck structure consists of three convolutional layers with kernel size of $1\times 1, 3\times 3$ and $1\times 1$. The first convolutional layer decreases channels by a ratio of $K$, thus the second convolutional layer with kernel size of $3\times 3$ has smaller input/output channel dimensions. Then the last convolutional layer restore channel dimensions. Many works have proven that bottleneck structure can efficiently decrease time complexity and model size in both CNN and FCN architecture \cite{he2016deep,vbnet}. 

Except bottleneck structure, we also adopt depth-wise separable convolutions (SepConv) which was presented in MobileNets \cite{howard2017mobilenets}. SepConv decompose a conventional convolution into a depth-wise convolution and a point-wise convolution. Generally, depth-wise convolutions aim to extract features with large kernel sizes. In contrast, point-wise convolutions aim to propagate the extracted features to required channel dimensions with small kernel size such as $1\times 1$. This scheme can efficiently decrease computation complexity and memory usage. 

\begin{figure}[bt]
    \centering
    \includegraphics[height=3cm]{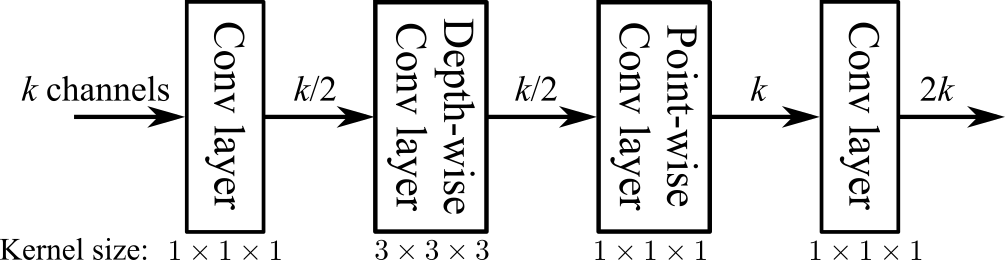}
    \caption{Compact ConvBlock structure.}
    \label{fig:convblock}
\end{figure}

By introducing bottleneck and SepConv structures, we present a modified convolution block, named \textit{Compact ConvBlock} (CCBlock), to compress the GPU memory demand while keeping same accuracy. The architecture is illustrated in Fig. \ref{fig:convblock}.

\subsection{Inplace-ABN and mixed-precision training}
Compact ConvBlock aims to decrease the number of weight parameters which leads to less memory requirements. To go one step further, we adopt Inplace-ABN layer \cite{rota2018place} to decrease the memory demand from implementation aspect. Modern deep learning frameworks such as Tensorflow \cite{tensorflow2015-whitepaper}, pyTorch \cite{paszke2017automatic} explicitly stores buffers after convolutional layer and normalization for the backward pass. Denote the gradient of the back-propagation through normalization as $\frac{\partial{L}}{\partial{x}}$, and gradient of the back-propagation through convolutional layer as $\frac{\partial{L}}{\partial{z}}$. $L$ denotes the loss, $x$ and $z$ are input feature maps to normalization and convolutional layer. In standard implementation, $x$ is buffered in GPU memory for backward pass. In Inplace-ABN implementation, $x$ in backward pass is approximated by cached $z$. Denote the approximation of $x$ as $\hat{x}$, then $\hat{x}$ can be described as: $\hat{x} = \dfrac{\phi^{-1}(z)-\beta}{\gamma}$, where $\phi(\cdot)$ denotes activation function. $\beta, \gamma$ denote the learnable parameters of batch normalization (BN) layer. By using the approximation, the buffer of $x$ can be released during training to save the run-time memory. For more details, please refer to original work \cite{rota2018place}. By using Inplace-ABN implementation, only buffer $z$ needs to be cached in GPUs instead of both $x$ and $z$. Up to $50\%$ of GPU memory can be saved theoretically.

In this work, we also try mixed-precision (MP) training technique \cite{micikevicius2018mixed} to further accelerate computation and decrease memory demand. Conventionally, we use single-precision format to allocate our data such as input training data, weight parameters, and gradients. A lot of researches has been conducted to explore advantages of half-precision format for deep learning \cite{courbariaux2014training,gupta2015deep}. However, pure half-precision training is exposed to underflow and overflow risk due to the narrow dynamic range. MP training presented several techniques to prevent the loss of critical information introduced by half-precision data. The key point of MP training is maintaining single-precision of weights that accumulates the gradients which are kept in half-precision format. To avoid loss of gradient information, loss scaling scheme is introduced to shift the gradient values into half-precision representable range. For more details please refer to Micikevicius \textit{et al.}'s paper\cite{micikevicius2018mixed}.

\section{Experiments} \label{sec:exp}
For experiments, we used a public dataset, \textit{KiTS19}, to validate our method. \textit{KiTS19} challenge aims to develop reliable kidney and kidney tumor semantic segmentation methods. The dataset contains 300 multi-phase CT images. 210 (70\%) of these cases are selected as training set. The dataset has a quite large slice pitch ranging from 0.5 mm to 5.0 mm. We apply same prepossessing as Isensee \textit{et al.} \cite{Isensee2019unet} who won the 1st place of the challenge. CT volumes are resampled to same voxel spacing of $3.22\times 1.62\times 1.62$ mm. The CT slice size of the resampled dataset ranges from 138 pixels to 329 pixels, number of slices ranges from 45 to 234. Median image shape is $128\times 248\times 248$ voxels. For each case, intensity is clipped by window width ranges from −79 to 304 H.U, and normalized by z-score normalization. 

Our networks were implemented with the PyTorch platform \cite{paszke2017automatic}. All experiments were performed on NVIDIA Tesla V100 GPU with 32 GB memory. We used official implementation of Inplace-ABN provided by Rota \textit{et al.}.  We used NVIDIA's API called automatic mixed precision (AMP) for MP training. Table \ref{table:gpu} demonstrate the GPU memory requirements of different architectures. We measured the actual GPU memory usage during training phase by the NVIDIA system management interface (nvidia-smi). We set batch size to one to measure the memory usage. Three different input sizes were experimented including two fixed patch sizes of $96\times 96\times 96$ and $256\times 256 \times 152$ voxels, and one dynamic size of $256\times 256 \times [45-234]$ voxels to cover full range of slices. Due to the limitation of our GPU memory, We need to limit our input slice size up to $256\times 256$ pixels. As demonstrated in Table \ref{table:gpu}, we can find that the CCBlock structure mainly decreased the number of parameters compared to the memory usage. In contrast, Inplace-ABN and MP training significantly decreased the memory usage up to 40\%. With the help of Inplace-ABN and MP training techniques, we can train the FCN using only one GPU card with 24 GB memory.

\section{Discussion} \label{sec:disc}

\begin{figure}[b]
    \centering
    \subcaptionbox{Input size: $96\times 96\times 96$ voxels\label{fig:curves_1}}{
        \includegraphics[width=0.45\textwidth,clip,trim=0 0 2cm 0cm]{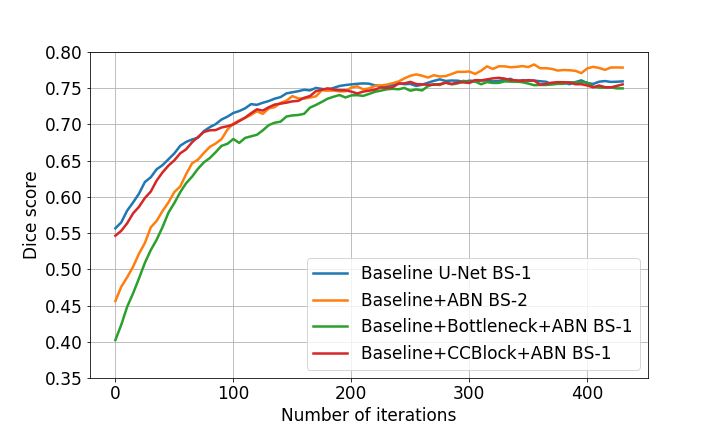}
    }
    \subcaptionbox{$256\times 256 \times 152$ voxels vs. $96\times 96\times 96$ voxels\label{fig:curves_2}}{
        \includegraphics[width=0.45\textwidth,clip,trim=0 0 2cm 0cm]{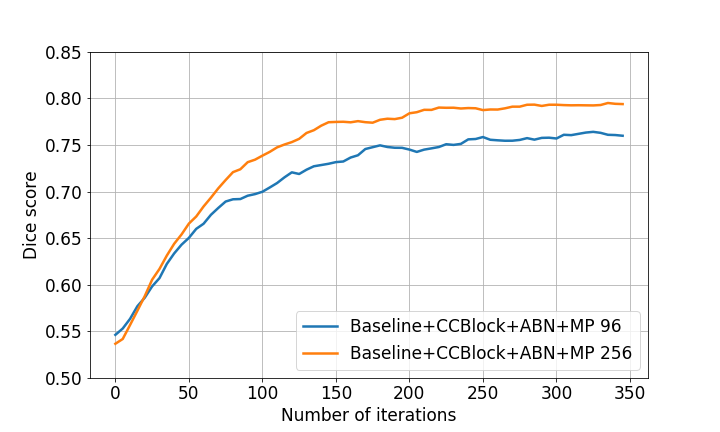}
    }
    \caption{Dice score of validations. (a) All validated networks were trained using subvolumes with size of $96\times 96\times 96$ voxels. (b) Same network with two different volume sizes. BS-$k$ denotes batch size of $k$.}
    \label{fig:curves}
\end{figure}

We validated the segmentation accuracy of different network architectures. All networks were trained using same hyper-parameter setting. We used a combination of Dice and cross-entropy loss \cite{Isensee2019unet}. Initial learning rate was set to $4e^{-4}$ with a learning rate scheduler that reduce the learning rate by 0.3 when training loss has stopped decreasing for 50 epochs. Adam optimizer was chosen to optimize networks. Validation curves are illustrated in Fig. \ref{fig:curves}. In Fig. \ref{fig:curves}(a), four different architectures were evaluated with same input size of $96\times 96\times 96$ voxels. Blue, yellow, green and red curves respectively denote baseline U-Net architecture with batch size of 1, baseline U-Net using Inplace-ABN structure with batch size of 2, baseline U-Net using bottleneck and Inplace-ABN structures with batch size of 1, and baseline U-Net with all auxiliary structures with batch size of 1. In Fig. \ref{fig:curves}(b), we validated same architecture with different input sizes. Architecture is baseline U-Net with all introduced methods including CCBlock, Inplace-ABN and MP training. Blue and yellow curves denote validation results with input sizes of $96\times 96\times 96$ voxels and $256\times 256\times 152$ voxels.

As demonstrated in Fig. \ref{fig:curves}(a), we can find that under the condition of same batch size, baseline U-Net and other variants had similar performance. The result demonstrates that our auxiliary modules can decrease the run-time memory and maintain the same accuracy. The yellow curve in Fig. \ref{fig:curves}(a) demonstrated the results of the training with batch size of two. Since we used batch normalization, experiments with two batch sizes slightly outperformed the one with single batch size. As future work, group normalization \cite{wu2018group} is considerable for single batch size training. From Fig. \ref{fig:curves}(b), we compared the influence of different input volume sizes. Experiment with the large input volume size outperformed the one with small size by 5\%. As previously mentioned, large FOV can contribute to a better learning performance. The experimental result has also proved the claim. In Fig. \ref{fig:compare}, we also demonstrate segmentation results of different sizes. As we can see, the segmentation results inferred by the model with large input size slightly outperformed the another model with smaller input crop size. 

\begin{figure}[b]
	\centering
     {\includegraphics[width=0.25\linewidth,trim={0 1cm 0 1cm},clip]{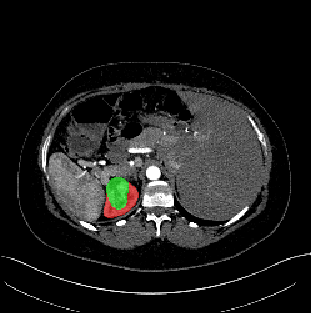}}\hspace{-1em}
     {\includegraphics[width=0.25\linewidth,trim={0 1cm 0 1cm},clip]{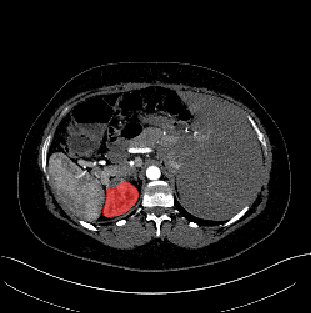}}\hspace{-1em}
     {\includegraphics[width=0.25\linewidth,trim={0 1cm 0 1cm},clip]{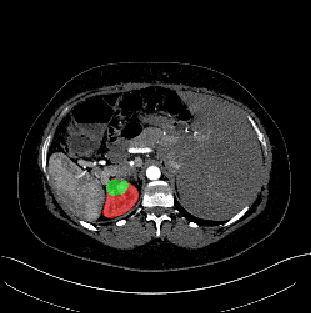}}\\
     
     {\includegraphics[width=0.25\linewidth,trim={0 1cm 0 0cm},clip]{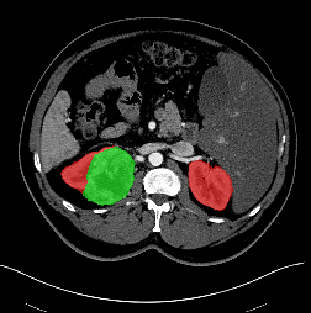}}\hspace{-1em}
     {\includegraphics[width=0.25\linewidth,trim={0 1cm 0 0cm},clip]{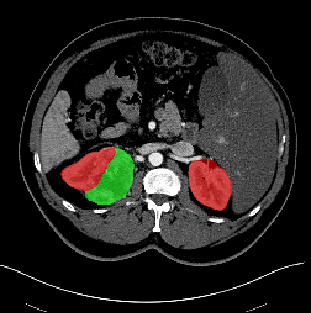}}\hspace{-1em}
     {\includegraphics[width=0.25\linewidth,trim={0 1cm 0 0cm},clip]{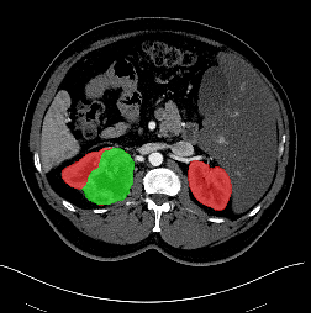}}\\
     
     \subcaptionbox{Ground truth}{\includegraphics[width=0.25\linewidth,trim={0 1cm 0 1cm},clip]{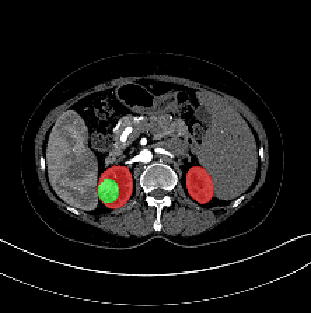}}\hspace{-1em}
     \subcaptionbox{$96\times 96\times 96$ voxels}{\includegraphics[width=0.25\linewidth,trim={0 1cm 0 1cm},clip]{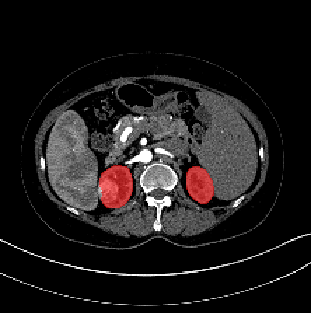}}\hspace{-1em}
     \subcaptionbox{$256\times 256 \times 152$ voxels}{\includegraphics[width=0.25\linewidth,trim={0 1cm 0 1cm},clip]{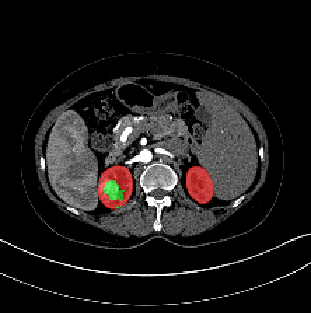}}\\
      \caption{Examples of segmentation results. (a) is ground truth labels. (b) and (c) are two comparison groups segmented by different models trained with different crop sizes. }
\end{figure}\label{fig:compare}

\begin{table}[]
\centering
\caption{Number of parameters and GPU memory usage of different networks.  Memory usage of two different input sizes is measured. Batch size was set to one for experiments.} \label{table:gpu}
\begin{adjustbox}{max width=\textwidth}
\begin{tabular}{lccccc}
\hline\noalign{\smallskip}
\multirow{2}{*}{} & \multirow{2}{*}{\textbf{Parameters \#}} & \multicolumn{3}{c}{\textbf{GPU Memory Usage} (per batch)} &  \\
&  & $96\times 96\times 96$ & $256\times 256 \times 152$ & $256\times 256 \times \text{full}$ &  \\ \hline
Baseline U-Net & 9.5 M & 3.7 GB & 28.4 GB  & $>$ 32 GB &  \\
Baseline+Bottleneck & 3.2 M & 3.1 GB & 28.3 GB  & $>$ 32 GB &  \\
Baseline+SepConv & 2.7 M & 3.2 GB &  30.1 GB & $>$ 32 GB &  \\
Baseline+CCBlock & 2.6 M & 3.0 GB & 27.2 GB & $>$ 32 GB &  \\
Baseline+CCBlock+Inplace-ABN & 2.6 M & 2.3 GB & 20.0 GB  & $\sim$ 30.2 GB & \\
Baseline+CCBlock+Inplace-ABN+MP & 2.6 M & 1.6 GB & 11.7 GB & $\sim$ 17.6 GB & \\
\hline\noalign{\smallskip}
\end{tabular}
\end{adjustbox}
\end{table}

\section{Conclusion}
Our presented memory-efficient FCN can reduce memory usage significantly. With less memory demand, we can feed larger volume to FCN to capture larger spatial context to achieve better segmentation performance. The proposed approach shows the potential of FCN in processing large medical images without subvolume cropping. In this work, we present a memory-efficient FCN for medical image segmentation tasks. We use bottleneck and depth-wise separable convolutions to slim our model, and use Inplace-ABN layer and mixed-precision training techniques to reduce the memory allocation in GPU. By adopting these techniques into our network, we saved up to 40\% GPU memory in our experiments. By reducing the memory usage, we can feed larger volumes to network even a whole volume without cropping, or feed more small subvolumes than conventional FCNs.

As for future works, the run-time memory usage still has room for improvement. In our experiments, to increase the input size as much as possible, we used small batch size of one or two which cannot contribute to batch normalization. As an alternative, in-place version of group or instance normalization may help this problem.

\section*{Acknowledgement}
Parts of this research was supported by MEXT/JSPS KAKENHI (26108006, 26560255, 17H00867, 17K20099), and the JSPS Bilateral Collaboration Grant and AMED (191k1010036h0001)


\bibliography{report}   
\bibliographystyle{spiebib}   

\end{document}